\documentclass[12pt]{article}
\usepackage{amsthm,amssymb,amsmath}
\usepackage[british]{babel}

 1 
1

\newcommand{\C}{\ensuremath{\mathbb{C}}}

\newcommand{\bt}{{\boldsymbol{t}}}
\newcommand{\ba}{{\boldsymbol{a}}}
\newcommand{\bT}{{\boldsymbol{T}}}

\renewcommand{\d}{\operatorname{d}}

\newcommand{\be}{\begin{equation}}
\newcommand{\ee}{\end{equation}}
\begin{document}

\title{\sc $\overline{\partial}$-approach to the dispersionless
KP hierarchy
\thanks{L. Martinez Alonso is partially supported by CICYT proyecto PB98--0821 }}
\author{B. Konopelchenko$^{1}$, L. Mart\'{\i}nez Alonso$^{2}$
 and  O. Ragnisco$^{3}$\\
\emph{ $^1$Dipartimento di Fisica, Universita
di Lecce}\\ \emph{73100 Lecce, Italy} \\
\emph{$^2$Departamento de F\'{\i}sica Te\'{o}rica II, Universidad
Complutense
}\\\emph{ E-28040 Madrid, Spain}\\
\emph{$^3$ Dipartimento di Fisica, Universita di Roma III}\\
\emph{Roma, Italy}
 }
\date{} \maketitle
\begin{abstract}

The dispersionless limit of the scalar nonlocal
$\overline{\partial}$-problem is derived. It is given by a special
class of nonlinear first-order equations. A quasi-classical
version of the $\overline{\partial}$-dressing method is presented.
It is shown that the algebraic formulation of dispersionless
hierarchies can be expressed in terms of properties of Beltrami
tupe equations. The universal Whitham hierarchy and, in
particular, the dispersionless KP hierarchy turn out to be  rings
of symmetries for the quasi-classical
$\overline{\partial}$-problem.

\end{abstract}

\vspace*{.5cm}

\begin{center}\begin{minipage}{12cm}
\emph{Key words:} Dispersionless KP hierarchy,
$\overline{\partial}$-equations.

\emph{ 1991 MSC:} 58B20.
\end{minipage}
\end{center}
\newpage

\section{Introduction}

Dispersionless or quasiclassical limits of integrable systems have
atracted a considerable interest during the last years (see e.g.
\cite{1}-\cite{13} and references therein). Such type of equations
and hierarchies arise as a result  of processes of averaging over
fast variables or as a formal quasiclassical limit $\hbar$(or
$\epsilon$) $\rightarrow 0$. Study of dispersionless hierarchies
is of great relevance since they play an important role in the
analysis of various problems in different fields of physics and
mathematics as, for example, the quantum theory of topological
fields and strings \cite{14}-\cite{16}, some models of optical
communications \cite{17} or the theory of conformal maps in the
complex plane \cite{18,19}.

Dispersionless hierarchies have been described and analysed by
different methods. In particular, the quasiclassical versions of
the inverse scattering transform and Riemann-Hilbert problem
method have been applied to the study of some $1+1$-dimensional
integrable equations \cite{23,20,11,12,13}. In contrast, similar
study of the $2+1$-dimensional integrable equations and
hierarchies is missing. Our goal is to fill this gap.

In the present paper we shall approach the dispersionless
hierarchies from the $\overline{\partial}$-dressing method. This
method, based on the linear nonlocal
$\overline{\partial}$-problem, is a very efficient tool for
constructing and solving usual integrable hierarchies (see e.g.
\cite{21,22,23}. We shall demonstrate that this approach provides
us with a new and promissing viewpoint of the dispersionless
hierarchies. First we shall derive the dispersionless (or
quasiclassical) version of the $\overline{\partial}$-problem for
the dKP hierarchy. It turns to be given by  nonlinear first-order
equations of the type
\begin{equation}\label{1.1}
\frac{\partial S}{\partial \bar{z}}=W\Big(z,\bar{z},\frac{\partial
S}{\partial z}\Big).
\end{equation}
It turns out that this type of equations are well-known in the
theory of complex analysis, in connection with quasi-conformal
mappings.

We shall formulate the quasi-classical version of the
$\overline{\partial}$-dressing method and derive the dKP hierarchy
using these equations. Moreover, we shall show that symmetries of
the quasiclassical $\overline{\partial}$-problem are determined by
a linear Beltrami equation and that they form an
infinite-dimensional ring structure which constitutes nothing but
the dKP hierarchy. In a more general setting this ring of
symmetries coincides with the universal Whitham hierarchy
introduced in \cite{8}.

\section{Dispersionless KP and universal Whitham hierarchies}

For the sake of convenience we remind here some relevant formulas
for the standard and dispersionless KP hierarchies. The standard
KP hierarchy written in Lax form (see e.g. \cite{7,10})
\begin{equation}\label{2.1}
\frac{\partial L}{\partial t_n}=[(L^n)_+,L],\quad n=1,2,\ldots,
\end{equation}
arises as the compatibility conditions for the system
\begin{equation}\label{2.2}
\begin{gathered}
L\psi=z\psi,\\
\frac{\partial \psi}{\partial t_n}=(L^n)_+\psi,\quad n=1,2,\ldots,
\end{gathered}
\end{equation}
where $L$ is a pseudo-differential operator
\[
L=\partial+u_1(\bt)\partial^{-1}+u_2(\bt)\partial^{-2}+\cdots,
\]
with
\[
\partial:=\frac{\partial}{\partial x},\quad \bt:=(t_1:=x,t_2,\ldots),
\]
$(L^n)_+$ denotes the pure differential part of $L^n$ and
$\psi=\psi(z,\bt)$ is a KP wave function.

The dKP hierarchy is given in Lax form by
\begin{equation}\label{d}
\frac{\partial \mathcal{L}}{\partial T_n}=\{(\mathcal{L}^n)_+,
\mathcal{L}\},\quad n=1,2,\ldots,
\end{equation}
where $\mathcal{L}=\mathcal{L}(p,\bT)$  denotes a function which
admits an expansion
\begin{equation}\label{z}
\mathcal{L}=p+\frac{U_1(\bT)}{p}+\frac{U_2(\bT)}{p^2}+\ldots,\quad
p\rightarrow\infty,\quad \bT:=(T_1:=X,T_2,\ldots),
\end{equation}
$(\mathcal{L}^n)_+$ is the polynomial part of $\mathcal{L}^n$ as a
function of $p$, and $\{,\}$ stands for the Poisson bracket
\[
\{f,g\}:=\frac{\partial f}{\partial p}\frac{\partial g}{\partial
x}-\frac{\partial f}{\partial x}\frac{\partial g}{\partial p}.
\]
This system of equations can be derived as a formal
$\epsilon\rightarrow 0$ limit of \eqref{2.1} under the change of
variables $T_n=\epsilon t_n,\; n\geq 1$ \cite{1}-\cite{10}. In
particular, the KP wave function is assumed to behave as
\[
\psi(z,\frac{\bT}{\epsilon})\sim \exp
\Big(\frac{S(z,\bT)}{\epsilon}+O(\epsilon ^0)\Big),
\]
where $S$ can be expanded as
\begin{equation}\label{s}
S(z,\bT)=\sum_{n\geq 1}z^n T_n+\sum_{n\geq
1}\frac{S_n(\bT)}{z^n},\quad z\rightarrow \infty,.
\end{equation}
Under these assumptions  it is easy to see that \eqref{2.2}
reduces to \eqref{z} and
\begin{equation}\label{2.6}
\frac{\partial S}{\partial T_n}=\Omega_n(p,\bT),
\end{equation}
where
\begin{equation}\label{2.66}
p:=\frac{\partial S}{\partial X},
\end{equation}
and  $\Omega(p,\bT):=(\mathcal{L}^n)_+$, with
$\mathcal{L}(p,\bT):=z(p,\bT)$ being the function provided by
solving for $z$ in \eqref{2.66}.

The system \eqref{2.6} represents a family of Hamilton-Jacobi
equations for $S$. Moreover, it can be shown \cite{10} that given
a function $S$ of the form \eqref{s} which satisfies \eqref{2.6},
then the corresponding function  $\mathcal{L}$ is a solution of
the dKP hierarchy \eqref{d}. The compatibility conditions for
\eqref{2.6} are given by
\begin{equation}\label{2.7}
\frac{\partial \Omega_m}{\partial T_m} -\frac{\partial
\Omega_n}{\partial T_n}+\{\Omega_n,\Omega_m\}=0,
\end{equation}
and represent the Zakharov-Shabat formulation of the dKP
hierarchy.

A general scheme for generating dispersionless hierarchies is the
universal Whitham hierarchy \cite{8}. Its starting point is a
Zakharov-Shabat system
\begin{equation}\label{2.8}
\frac{\partial \Omega_A}{\partial T_B} -\frac{\partial
\Omega_B}{\partial T_A}+\{\Omega_A,\Omega_B\}=0,
\end{equation}
with  $\Omega_A=\Omega_A(p,\bT)$ being given  meromorphic
functions of $p$ depending on a set of parameters $\bT$. This
hierarchy includes as particular cases the dKP, Benney and the
dispersionless version of the $2$-dimensional Toda lattice.

\section{Quasi-classical $\bar{\partial}$-problems}

The standard KP hierarchy is associated with the following scalar
non-local linear $\bar{\partial}$ equation (see e.g.
\cite{21}-\cite{23}) for the KP wave function
\begin{equation}\label{3.1}
\frac{\partial \chi(z,\bar{z},\bt)}{\partial \bar{z}}=
\int\!\!\int_G \d z'\d \bar{z}'
\chi(z',\bar{z}',\bt)\psi_0(z',\bt)R_0(z',\bar{z}',z,\bar{z})\psi_0^{-1}(z,\bt),
\end{equation}
where $G$ is a given bounded domain of $\C$, $
\psi_0(z,\bt)=\exp(\sum_{n\geq 1} z^n t_n)$
 and $R_0=
R_0(z',\bar{z}',z,\bar{z})$ an appropriate function
($\bar{\partial}$-data). It is assumed that the function $\chi$
has a canonical normalization
\[
\chi(z,\bar{z},\bt)=1+\frac{\chi_1(\bt)}{z}+\frac{\chi_2(\bt)}{z^2}+\ldots,\quad
z\rightarrow\infty.
\]
The corresponding wave function of the standard KP hierarchy is
then given by $\psi=\psi_0\cdot\chi$.

From \eqref{3.1} it follows that in order to get a non singular
dispersionless limit of the corresponding solution of the KP
hierarchy, the function
\begin{equation}\label{3.3}
\begin{gathered}
u(\frac{\bT}{\epsilon}):= \epsilon\frac{\partial }{\partial T_1}
\int\!\!\int_G \d z\d \bar{z}\int\!\!\int_G \d z'\d \bar{z}'R_0(z',\bar{z'},z,\bar{z}) \\
\times\exp\Big(\frac{1}{\epsilon}(S_0(z',\bT)-S_0(z,\bT))\Big) ,
\quad S_0(z,\bT):=\sum_{n\geq 1} z^nT_n,
\end{gathered}
\end{equation}
should have a finite $\epsilon\rightarrow 0$ limit. This holds, in
particular, for $\bar{\partial}$-data of the form
\begin{equation}\label{3.4}
R_0(z',\bar{z'},z,\bar{z})=\sum_{k\geq 0}(-1)^k r_k(z,\bar{z}')
\epsilon^{k-1}\delta^{(k)}(z'-z-\epsilon\alpha_k(z,\bar{z})),
\end{equation}
where $r_k$ and $\alpha_k$ are arbitrary functions and
\[
\delta^{(k)}(z,\bar{z}):=\frac{\partial^k
\delta(z,\bar{z})}{\partial z^k}.
\]

If we now rewrite \eqref{3.1} as
\begin{equation}\label{3.5}
\frac{\partial \ln\chi(z,\bar{z},\frac{\bT}{\epsilon})}{\partial
\bar{z}}= \int\!\!\int_G \d z'\d \bar{z}'
\psi(z',\bar{z'},\frac{\bT}{\epsilon})
R_0(z',\bar{z'},z,\bar{z})\psi(z,\bar{z},\frac{\bT}{\epsilon})^{-1},
\end{equation}
and insert a kernel of the form \eqref{3.5}, the limit
$\epsilon\rightarrow 0$ leads to
\begin{equation}\label{3.8}
\frac{\partial S}{\partial \bar{z}}=W(z,\bar{z},\frac{\partial
S}{\partial z}),
\end{equation}
where
\begin{equation}\label{3.7}
W(z,\bar{z},\frac{\partial S}{\partial z}):=\sum_{k\geq
0}r_k(z,\bar{z})\exp\Big(\alpha_k(z,\bar{z})\frac{\partial
S}{\partial z}\Big) \Big(\frac{\partial S}{\partial z}\Big)^k.
\end{equation}

The above discussion suggests to take equation \eqref{3.8}, for
appropriate functions $W$,  as the quasi-classical version of the
linear $\bar{\partial}$-problem \eqref{3.1}. The function $S$ is
widely used in the discussions of the dispersionless limit of the
integrable hierarchies \cite{4}-\cite{10}. Within the
$\bar{\partial}$-approach it is a non-holomorphic function of the
spectral parameter and obeys the nonlinear
$\bar{\partial}$-equation \eqref{3.8}.

The nonlinear Beltrami type equation \eqref{3.8} is well-known in
complex analysis. Under certain conditions on $W$ it belongs to
the class of  nonlinear elliptic systems on the plane  in the
sense of Lavrent'ev \cite{24, 27,28}. On the other hand, solutions
of equations of this type determine quasi-conformal maps of their
domains of definition  on the complex plane ( see e.g. \cite{24,
25}). The connection between dispersionless hierarchies and the
theory of quasiconformal maps is an interesting problem which will
be considered elsewhere.

\section{Quasi-classical $\bar{\partial}$-dressing method}

Now we will use the $\bar{\partial}$-problem \eqref{3.8} to
formulate the dKP hierarchy. In what follows we will have in mind
expressions for $W$ in which the dependence on $z$ and $\bar{z}$
are provided for compact supported functions so as to allow for
solutions of \eqref{3.8} with asymptotic form \eqref{s}.

 Suppose given a solution
$S=S(z,\bar{z},\bT)$ of \eqref{3.8} which as $z\rightarrow\infty$
is of the form \eqref{z}. Then
\begin{equation}\label{4.2}
\frac{\partial }{\partial \bar{z}}\Big( \frac{\partial S}{\partial
T_n}\Big)=W'\Big(z,\bar{z},\frac{\partial S}{\partial z}\Big)
\frac{\partial }{\partial z}\Big( \frac{\partial S}{\partial
T_n}\Big),\quad n\geq 1,
\end{equation}
where
\begin{equation}
W'(z,\bar{z},\lambda):=\frac{\partial
W(z,\bar{z},\lambda)}{\partial \lambda}.
\end{equation}
This means that all time derivatives of $S$ satisfy the same
family (dependent on the infinite set of parameters $\bT$) of
linear Beltrami equations
\begin{equation}\label{Beltrami}
\frac{\partial \Phi}{\partial \bar{z}}=Q(z,\bar{z},\bT)
\frac{\partial \Phi}{\partial z},
\end{equation}
where
\[
Q(z,\bar{z},\bT):=W'\Big(z,\bar{z},\frac{\partial S}{\partial
z}\Big).
\]
Together with $\frac{\partial S}{\partial T_n}$, any combination
\[
\sum_{k} U_{n_k}(\bT)\Big(\frac{\partial S}{\partial
T_{n_k}}\Big)^{m_k},
\]
obeys \eqref{4.2} as well. On the other hand, under mild
conditions \cite{25,26}, a solution of the linear Beltrami
equation bounded on the whole plane $\C$ and vanishing at
$z\rightarrow\infty$ vanishes identically. These properties are
fundamental for our formulation of the dKP hierarchy. Indeed,
given a solution of \eqref{3.8} of the form \eqref{z}  then it
follows that
\begin{equation}\label{ss1}
\frac{\partial
S}{\partial T_n}=z^n+\sum_{m\geq 1}\frac{S_m(\bT)}{z^m},\quad
z\rightarrow\infty,
\end{equation}
and, in particular,  the function $p:=\frac{\partial S}{\partial
X}$ can be expanded as
\begin{equation}\label{p}
p=z+\sum_{n\geq 1}\frac{\partial_X S_n}{z^n},\quad
z\rightarrow\infty.
\end{equation}
In this way, if we denote by $\mathcal{L}(p,\bT)$ the expansion
for $z$ obtained by inverting  \eqref{p}, it is clear from
\eqref{ss1} and \eqref{p} that
\[
\frac{\partial S}{\partial
T_n}-(\mathcal{L}^n)_+=O(\frac{1}{z}),\quad z\rightarrow\infty.
\]
Hence, as
\[
\frac{\partial S}{\partial T_n}-(\mathcal{L}^n)_+
\]
is also a solution of \eqref{Beltrami} we conclude that $S$
satisfies \eqref{2.6}, so that $\mathcal{L}(p,\bT)$ is a solution
of the dKP hierarchy. Therefore, we see that the quasi-classical
$\bar{\partial}$-problem \eqref{3.8} leads to a straightforward
formulation of the dKP hierarchy along the standard logic of the
$\bar{\partial}$-dressing method.

\section{Ring of symmetries of the quasi-classical
$\bar{\partial}$-problem and the universal Whitham hierarchy}

In section 3 we have derived the $\bar{\partial}$-problem
\eqref{3.8} by starting with the $\bar{\partial}$-problem
\eqref{3.1} for the KP hierarchy. One can show that the
quasi-classical cs of the form \eqref{3.8} arise also as the
dispersionless limit of other scalar integrable hierarchies, like
the $2$-dimensional Toda lattice and the modified KP hierarchy.
The only difference consists in the different behaviours assumed
for $S$ at infinity.

 Thus the problem \eqref{3.8}, taken on some bounded
domain $G$ of $\C$, can be regarded as the starting point of a
whole approach to scalar dispersionless hierarchies without any
reference to the linear $\bar{\partial}$-problems of the original
\emph{dispersionfull} hierarchies. The main feature of this
approach is that the symmetries (first order variations $\delta
S$) of \eqref{3.8} are determined by the Beltrami equation
\begin{equation}\label{5.2}
\frac{\partial }{\partial \bar{z}}(\delta S)
=W'\Big(z,\bar{z},\frac{\partial S}{\partial z}\Big)
\frac{\partial }{\partial z}(\delta S).
\end{equation}
As a consequence, if $\delta S$ is a solution of \eqref{5.2} and
$\Phi(\xi)$ is a differentiable function, then $\Phi(\delta S)$
is a solution too. Reciprocally, given two solutions $\delta _i S$
$(i=1,2)$ there exists a function $\Phi(\xi)$ such that $\delta_1
S=\Phi(\delta_2 S)$. Moreover, the product of two symmetries
$\delta_1 S\delta_2 S$ also satisfies \eqref{5.2}.

 Therefore, symmetries of the quasi-classical
$\bar{\partial}$-problem \eqref{3.8} form an infinite-dimensional
ring.

Let us denote by $T_A$ the times associated to the corresponding
symmetry flows, so that the infinitesimal symmetries are
$\frac{\partial S}{\partial T_A}$. If we mark one of such
symmetries $\frac{\partial S}{\partial T_{A_0}}$ and denote it by
$p$, then for any symmetry we can write
\begin{equation}\label{5.3}
\frac{\partial S}{\partial T_A}=\Omega_A(p,\bT),
\end{equation}
for a certain function $\Omega_A$. Thus, by varying $\Omega_A$ one
can generate all the symmetries of the $\bar{\partial}$-problem
\eqref{3.8} out of one of them $p$ (basically arbitrary). The
compatibility conditions for \eqref{5.3} are
\begin{equation}\label{5.4}
\frac{\partial \Omega_A}{\partial T_B} -\frac{\partial
\Omega_B}{\partial T_A}+\{\Omega_A,\Omega_B\}=0,
\end{equation}
where
\[
\{f,g\}:=\frac{\partial f}{\partial p}\frac{\partial g}{\partial
A_0}-\frac{\partial f}{\partial A_0}\frac{\partial g}{\partial p}.
\]

The infinite system \eqref{5.4} is exactly the Universal Whitham
hierarchy. Hence, this hierachy describes the infinite-dimensional
ring of symmetries of the scalar quasi-classical
$\bar{\partial}$-problem \eqref{3.8}. As it was shown in \cite{8}
the Universal Whitham hierarchy contains several relevant
dispersionless hierarchies as particular reductions. This means
that the quasi-classical $\bar{\partial}$-problem \eqref{3.8} has
also a universal character.

\section{Solutions of dispersionless hierarchies }

Quasi-classical $\bar{\partial}$-dressing methods based on
\eqref{3.8} provide us also with a tool for solving dispersionless
hierarchies. The point is that we can apply the method of
characteristics to solve \eqref{3.8} and then  to find solutions
$S$ satisfying the appropriate behaviour at $\infty$.

To illustrate this process let us consider the dKP hierarchy and a
$\bar{\partial}$-problem of the form
\begin{equation}\label{dbar}
\frac{\partial S}{\partial
\bar{z}}=\theta(1-z\bar{z})W_0\Big(\frac{\partial S}{\partial
z}\Big ),
\end{equation}
where $\theta(\xi)$ is the usual Heaviside function and $W_0(m)$
is an arbitrary differentiable function.  Observe that
\eqref{dbar} implies
\[
\frac{\partial m}{\partial \bar{z}}=W'_0(m)\frac{\partial m
}{\partial z},\quad m:=\frac{\partial S}{\partial z}, \quad |z|<1,
\]
where $W'_0=\frac{\d W_0}{\d m}$. This equation can be solved at
once by applying the methods of characteristics, so that the
general solution $S_{in}$ of \eqref{dbar} inside the unit circle
$|z|<1$ is implicitly characterized by
\begin{equation}\label{gen}
\begin{gathered}
S_{in}=W_0(m)\bar{z}+mz-f(m),\\ \\W'_0(m)\bar{z}+z=f'(m),
\end{gathered}
\end{equation}
where $f=f(m)$ is an arbitrary differentiable function. The
solution $S_{out}$ of \eqref{dbar} outside the unit circle is any
arbitrary analytic function ($\bar{z}$-independent). However, in
order to obtain a global solution of \eqref{dbar} in the class of
locally integrable generalized functions we impose the continuity
of $S$ at the unit circle, so that
\begin{equation}\label{out}
S_{out}(z)=S_{in}(z,\frac{1}{z}),\quad |z|=1.
\end{equation}
Moreover,  as we are dealing with the dKP hierarchy, we require
$S_{out}$ to be of the form
\begin{equation}\label{ss1}
S_{out}= \sum_{n\geq 1}z^n T_n+\sum_{n\geq 1}\frac{s_n(\bT)}{z^n}.
\end{equation}
On the other hand, from \eqref{gen} it follows that the
coefficients of $S_{out}$ are determined by the identity
\begin{equation}\label{eq1}
\frac{\partial S_{out}}{\partial z}=m_0-\frac{W(m_0)}{z^2},\quad
m_0(z):=m(z,\frac{1}{z}),
\end{equation}
where $m_0$ is related to  the arbitrary function $f$ in
\eqref{gen} as
\begin{equation}\label{eq1}
\frac{W'(m_0)}{z}+z=f'(m_0).
\end{equation}

In  this way we have a solution method for the dKP hierarchy based
on solving the $\bar{\partial}$-equation \eqref{dbar}. In
principle  the process is the following. We first take a certain
function $f=f(m,\ba)$ depending on $m$ and a certain set of
undetermined parameters $\ba=(a_1,\ldots,a_n)$, and solve for
$m=m(z,\bar{z},\ba)$ in the second equation of \eqref{gen}. Then
the functions $S_{in}(z,\bar{z},\ba)$ and $S_{out}(z,\ba)$ are
determined by means of the first equation of \eqref{gen} and
\eqref{out}, respectively. Finally, we impose $S_{out}$ to admit
an asymptotic expansion \eqref{ss1} and get the parameters $\ba$
as functions of the dKP times $\bt$. However, as this method of
solution is based on solving implicit equations, there is no
general guarantee that the  resulting expression for  $S$ be a
global generalized solution of \eqref{dbar}, so that a further
analysis of the regularity of $S$  is required. Furthermore, in
case we were able to guarantee the appropriate  regularity of $S$
there would be an easier way to determine  $S$. Indeed, we may
start with a function $m_0=m_0(z,\ba)$ and find the coefficients
of the expansion \eqref{ss1} of $S_{out}$ by imposing \eqref{ss1}
and \eqref{eq1}. According to \eqref{eq1}, the corresponding
function $f=f(m,\ba)$ in \eqref{gen} is  then given by
\[
f'(m_0,\ba)=\frac{W'(m_0)}{h(m_0,\ba)}+h(m_0,\ba),
\]
where $h(m_0,\ba)$ is the inverse function  $z=h(m_0,\ba)$ of the
function $m_0=m_0(z,\ba)$.

  As an example, let us consider the
case
\[
W_0(m)=m^2,\quad m_0:=az^2+bz+c+\frac{d}{z}.
\]
From \eqref{eq1} we get a function $S_{out}$ of the form
\[
S_{out}=\sum_{n=1}^3 z^n T_n+\sum_{n=1}^3\frac{S_n(\bT)}{z^n},
\]
where
\[
\begin{gathered}
a-a^2=3T_3,\quad (1-2a)b=2T_2,\\ \\
(1-2a)c-b^2=X,\quad (1-2a)d-2bc=0,\\\\
S_1=c^2+2bd,\quad S_2=cd,\quad S_3=\frac{d^2}{3}.
\end{gathered}
\]
We notice that the function
\[
U:=-\frac{\partial S_1}{\partial X}=-\frac{2X+24T_2^2-24T_3
X}{(1-12T_3)^2},
\]
verifies the first equation of the dKP hierarchy
\[
\frac{3}{2}\frac{\partial^2 U}{\partial T_2^2}=\frac{\partial
}{\partial X}\Big( \frac{\partial U}{\partial
T_3}-6U\frac{\partial U}{\partial X}\Big).
\]

A general discussion of the $\bar{\partial}$-method of solution
for dispersionless hierarchies will be presented elsewhere.


\newpage


\begin{thebibliography}{99}
\bibitem{1} D. Lebedev and Y. Manin, Phys. Lett. \emph{74 A},
154-156 (1979)
\bibitem{2} V. E. Zakharov, Func. Anal. Priloz. \emph{14}, 89-98
(1980); Physica \emph{3D}, 193-202 (1981)
\bibitem{3} P. D. Lax and C. D. Levermore, Commun. Pure Appl. Math., \emph{36}, 253-290, 571-593, 809-830 (1983   )
\bibitem{4} I. M. Krichever, Func. Anal. Priloz., \emph{22}, 37-52  (1988)
\bibitem{5} Y. Kodama, Phys. Lett. \emph{129A}, 223-226  (1988)
\bibitem{6} B. A. Dubrovin and S. P. Novikov, Russian Math. Surveys, \emph{44}, 35-124 (1989)  )
\bibitem{7} K. Takasaki and T. Takebe, Int. J. Mod. Phys. A, Suppl.\emph{1B}, 889-922 (1992)
\bibitem{8} I. M. Krichever, Commun. Pure Appl. Math., \emph{47}, 437-475 (1994)
\bibitem{9} \emph{Singular limits of dispersive waves} (eds. N. M. Ercolani et al), Nato  Adv. Sci.
Inst. Ser. B Phys. \emph{320 }, Plenum, New York  (1994)
\bibitem{10} K. Takasasi and T. Takebe, Rev. Math. Phys., \emph{7}, 743-808 (1995)
\bibitem{11} R. Carrol and Y. Kodama,  J. Phys. A Math. Gen. \emph{28}, 6373-6387 (1995)
\bibitem{12} S. Jin, C. D. Levermore and D. W. McLaughlin, Comm. Pure and Appl. Math.,
 \emph{52}, 613-654 (1999)
\bibitem{13} S. Kamvissis, K. T-R McLaughlin and P. D. Miller, \emph{Semiclassical soliton emsembles
for the focusing nonlinear Schr \"{o}dinger equation }, nlin.
SI/0012034 (2000)
\bibitem{14} I. M. Krichever, Comm. Math. Phys., \emph{143}, 415-429  (1992)
\bibitem{15} B. A. Dubrovin, Comm. Math. Phys. \emph{145}, 195-207  (1992)
\bibitem{16} S. Aoyama and Y. Kodama, Commun. Math. Phys. \emph{182}, 185-219  (1996)
\bibitem{17} Y. Kodama, \emph{The whitham equations for optical communications: mathematical theory
of NR2},  solv-int/9709012 (1997)
\bibitem{18} P. B. Wiegmann and A. Zabrodin, Commun. Math. Phys \emph{213}, 523-538 (2000)
\bibitem{19} M. Mineev-Weinstein P. B. Wiegmann and A. Zabrodin, Phys. Rev. Lett. \emph{84},
5106-5109  (2000)
\bibitem{20} V. V. Geogdzhaev, Teor. Mat. Fiz. \emph{73},  255-263 (1987)
\bibitem{21} V. E. Zakharov and S. V. Manakov, Func. Anal. Appl. \emph{19}, 89-101  (1985)
\bibitem{22} V. E. Zakharov, \emph{On the inverse method} in \emph{Inverse Problems in Action}
(P. S. Sabatier, Ed), Springer-Verlag, Berlin (1990)
\bibitem{23} B. G. Konopelchenko, \emph{Solitons in multidimensions}, World Scientific, Singapore
 (1993)
\bibitem{24} L. V. Ahlfors, \emph{Lectures on quasi-conformal mappings}, D. Van Nostrand C.
, Princeton  (1966)
\bibitem{25} L. Bers, Bull. American Math. Soc, \emph{83}, 1083-1100  (1977)

\bibitem{26} I. N. Vekua, \emph{Generalized analytic functions}, Pergamon Press, Oxford  (1962)


\end{thebibliography}
\end{document}